\documentclass[aps,prl,amsmath,amssymb,twocolumn,floatfix,10pt,showpacs]{revtex4-1}
\usepackage{graphicx}
\usepackage{hyperref}

\newcommand{\bs}{\;\;\;\;\;}
\newcommand{\sms}{\;\;}
\newcommand{\ve}{\mathbf}

\begin{document}

\title{Helical modes in carbon nanotubes generated by strong electric fields}
\author{Jelena Klinovaja}
\author{Manuel J. Schmidt}
\author{Bernd Braunecker}
\author{Daniel Loss}
\affiliation{Department of Physics, University of Basel, Klingelbergstrasse 82, 4056 Basel, Switzerland}
\date{\today}


\begin{abstract}
Helical modes, conducting opposite spins in opposite directions, are shown to exist in metallic armchair nanotubes in an all-electric setup. This is a consequence of the interplay between spin-orbit interaction and strong electric fields. The helical regime can also be obtained in chiral metallic nanotubes by applying an additional magnetic field.
In particular, it is possible to obtain helical modes at one of the two Dirac points only, while the other one remains gapped. Starting from a tight-binding model we derive the effective low-energy Hamiltonian and the resulting spectrum.
\end{abstract}

\maketitle

Carbon based solid state physics has attracted much attention over the past decades. One of the best studied structures in this field is the carbon nanotube (CNT), a hexagonal lattice of carbon atoms rolled up to a cylinder \cite{dresselhaus_book}. The experimental techniques for creating, isolating, and analyzing CNTs have by now remarkably matured, such that characteristics that have previously been obscured by disorder can now be experimentally resolved \cite{kuemmeth_nature, marcus_kuemmeth_nature,kouwenhoven_steele, schoenenberger}. An example is the spin-orbit interaction (SOI), which is generally small in CNTs \cite{ando,huertas-hernando,jeong,izumida}, yet can affect electron spin decoherence in CNT quantum dots \cite{bulaev,rudner_rashba}, or allow spin control \cite{flensberg_marcus,burkard} and spin filtering \cite{spin_peierls}. A complete understanding of the SOI in CNTs becomes therefore desirable.

In this Letter, we investigate the effect of SOI in combination with a strong electric field in single-wall CNTs within an effective low-energy theory. In particular, we
identify experimentally accessible parameter regimes in which SOI and electric fields create helical modes without the need for magnetic fields. 
This must be contrasted with the helical modes in one-dimensional metals with Rashba SOI, which can be created only with an additional magnetic field that opens a gap at the crossing point of the two Rashba-shifted parabolas \cite{spin_peierls,wires}. 
Helical modes, conduction channels transporting opposite spins in opposite directions, naturally lead to spin filtering, but they have also potential application as Cooper pair splitters \cite{tserkovnyak} and, if in proximity with a superconductor, lead to Majorana bound states at the edges of the conductor \cite{majorana}.
Helical modes have also attracted much attention recently in the context of topological insulators \cite{hasan_kane}.
Such physics may be achieved in CNTs in an all-electric setup.

Perfect helical modes appear in armchair CNTs, while in metallic chiral CNTs the spins of the left and right moving modes are not precisely opposite. In the latter, however, perfect helicity can be restored in one Dirac point by an additional magnetic field, whereas the other Dirac point becomes insulating at these energies. This corresponds to the effective suppression of one valley for the low-energy physics.

\begin{figure}[!ht]
\centering
\includegraphics[width=150pt]{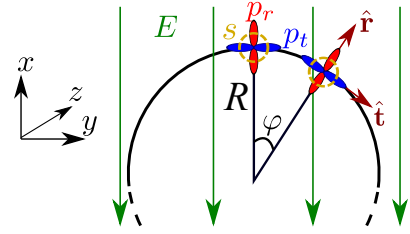}
\caption{(Color online) Cross section of a CNT in a uniform electric field $E$. The orientation of the orbitals $p_r,p_t$ as well as the local coordinate system $\hat {\ve r},\hat{\ve t}$ depends on the azimuthal angle $\varphi$. The $s$ orbital is indicated by the dashed circles. The electric field $E$ is oriented along the $x$-direction of the global coordinate system. The $z$-direction is along the nanotube.}
\vskip-7pt
\label{fig_tube_with_field}
\end{figure}

{\it The model.} The effective theory is based on a comprehensive model which incorporates the curvature effects for nearest-neighbor hopping and orthogonal orbitals \cite{dresselhaus_book}.
Charge effects in CNTs due to electric fields have been considered before \cite{benedict_screening, novikov_levitov, dft_screening}. Here, we also include spin effects induced by external uniform electric
fields (see Fig. \ref{fig_tube_with_field}). For this we start from a tight-binding description of the honeycomb lattice on a cylinder surface where we include all orbitals of the second
shell and the hybridization of the $\pi$ and the $\sigma$ bands.
The screening of the electric field by electron-electron interactions is treated on the mean-field level. The corresponding Hamiltonian is
\begin{equation}
H = H_{\rm bs} + H_{\rm SO} + H_{E}^{(1)} + H_{E}^{(2)}.
\label{full_hamiltonian}
\end{equation}
The band structure Hamiltonian $H_{\rm bs}$ includes the hopping of electrons between orbitals of neighboring carbon atoms and accounts for the orbital energies $H_{\rm bs} = t_{ij}^{\mu\mu'} c^\dagger_{i\mu\lambda} c_{j\mu'\lambda} + \varepsilon_s c^\dagger_{is\lambda} c_{is\lambda}$. Here $c_{i\mu\lambda}$ are the electron operators, $i$ and $j$ are nearest neighbor sites on the honeycomb lattice, $\lambda = \pm 1$ is the spin in $z$-direction, and $\mu$ runs over the second shell orbitals with $\mu=s$ the $s$ orbital and $\mu=p_r,p_t,p_z$ the $p$ orbitals pointing in radial, tangential and $z$-direction (see Fig. \ref{fig_tube_with_field}). The $\pi$ band is formed by the $p_r$ orbitals, while the $\sigma$ band is formed by $p_t,p_z,s$. Summation over repeated indices is assumed. The hopping amplitude $t^{\mu\mu'}_{ij}$ between $(j,\mu')$ and $(i,\mu)$ is a linear combination of the four fundamental hopping amplitudes 
$V_{ss},V_{sp},V_{pp}^\pi,V_{pp}^\sigma$ \cite{dresselhaus_book} with coefficients depending on the relative orientation of the orbitals $\mu$ and $\mu'$ \cite{schmidt_loss_GG_interfaces}. The energy difference between $s$ and $p$ orbitals is $\varepsilon_s$.

The atomic SOI is modeled by the on-site Hamiltonian $H_{\rm SO} = i\Delta_{\rm SO} \varepsilon^{\mu\nu\eta} c^\dagger_{i\mu\lambda} S^\nu_{\lambda\lambda'} c_{i\eta\lambda'}$, where now $\mu,\eta=p_r,p_t,p_z$, $\varepsilon^{\mu\nu\eta}$ is the Levi-Civita symbol, and $\Delta_{\rm SO} = 6$meV \cite{min_soi}. The index $\nu=r,t,z$ labels the spin components in the local coordinate system, i.e., $S^r = S^x \cos\varphi_i  + S^y\sin\varphi_i$, $S^t = S^y \cos\varphi_i - S^x \sin\varphi_i$, with $\varphi_i$ the azimuthal angle of site $i$ (see Fig. \ref{fig_tube_with_field}) and $S^{x,y,z}$ the spin Pauli matrices (with eigenvalues $\pm 1$). 

An electric field oriented perpendicular to the tube axis affects the electrons in two ways. First, the orbital energies are modulated by the electrostatic potential gradient. This is described by the on-site energy Hamiltonian $H_E^{(1)} = e E^* R \cos(\varphi_i) c^\dagger_{i\mu\lambda} c_{i\mu\lambda}$, where $E^*$ is the screened electric field, $e$ is the electron charge, and $R$ is the CNT radius. This Hamiltonian induces a rearrangement of charges on the CNT surface and so, by Coulomb interaction, leads to screening of $E$. Hence, $H_E^{(1)}$ depends on the screened field inside the tube $E^*$, which we find in the linear regime to be given by $E^* = E/\gamma$ with $\gamma\simeq 5$, in agreement with Refs. \cite{benedict_screening,novikov_levitov,dft_screening}. However,  the renormalization of the Fermi velocity $v_F$ \cite{novikov_levitov} is found to be negligible for the parameters used in this paper.

Second, an electrostatic potential $\phi(\ve r)$ varying on the lattice scale induces intra-atomic transitions between orbitals $\mu$ and $\mu'$ because generally $\left<\mu\right| \phi(\ve r)\left|\mu'\right>\neq 0$. Most important is the $s$-$p_r$ transition because of two reasons: 1) It is the only transition directly coupling $\pi$ and $\sigma$ bands, thus giving rise to a first order effect in the $s$-$p_r$ coupling strength. 2) Its strength is determined by the unscreened field $E$ and not by $E^*<E$. Indeed, the induced potential $\phi_{\rm ind}$ cancels in $H_E^{(2)}$, i.e. $\left<p_r\right|\phi_{\rm ind}(r)\left|s\right>=0$, as $\phi_{\rm ind}$ is approximately an even function in $r$ about $r=R$. Based on these arguments, we keep only the $s$-$p_r$ transition. The validity of this approximation was also verified numerically. The resulting Hamiltonian is $H_{E}^{(2)} = - eE \xi_0 \cos(\varphi_i) c^\dagger_{ip_r\lambda} c_{is\lambda} + {\rm H.c.}$, where $\xi_0 = -\left<p_r\right| r \left|s\right>=\frac{3a_B}Z\simeq 0.5${\AA} with $a_B$ the Bohr radius and $Z\simeq 3.2$, where we have assumed hydrogenic wave functions for the second shell carbon orbitals.

\begin{table*}[hbt]
\caption{The effective Hamiltonian for CNTs. $a \simeq 2.4$\AA\, is the lattice constant. $\theta$ is the chiral angle ($\theta=\frac{\pi}{6}$ for armchair CNTs). $\sigma_{1,2}$ are the Pauli matrices in sublattice space. $S^{x,y,z}$ are the spin operators (eigenvalues $\pm1$).
$(V_{ss},V_{sp}, V_{pp}^\pi, V_{pp}^\sigma, \varepsilon_s) = -(6.8,\,5.6,\,3.0,\,5.0,\,8.9)$ eV \cite{dresselhaus_book}, $\Delta_{\rm SO} = 6$ meV \cite{min_soi}. The Fermi velocity is $v_F = \sqrt3 |V_{pp}^\pi|a/2\hbar \simeq 0.95 \times 10^6$ m/s.}
\begin{tabular}{ll}
\hline\hline \\[-10pt]
$H_{\rm orb}^{\rm cv} = \hbar v_F (\Delta k_{\rm cv}^t \sigma_1 + \tau\Delta k_{\rm cv}^z \sigma_2)\,\,^{\rm *)}\sms$  
& 
$\hbar v_F \Delta \ve k_{\rm cv} = \hbar v_F \begin{pmatrix}\Delta k_{\rm cv}^t\\\Delta k_{\rm cv}^z\end{pmatrix} = \tau \dfrac{V^{\pi}_{pp}(V^{\pi}_{pp}-V^{\sigma}_{pp})}{8 (V^{\pi}_{pp}+V^{\sigma}_{pp})}\left(\dfrac{a}{R}\right)^2 \begin{pmatrix} -\cos 3\theta\\\sin 3\theta\end{pmatrix} \simeq  \tau \dfrac{5.4\, {\rm meV}}{R[{\rm nm}]^2}\begin{pmatrix} -\cos 3\theta\\\sin 3\theta\end{pmatrix}$ \\[10pt]
\hline\\[-10pt]
$H_{\rm SO}^{\rm cv}= \alpha S^z\sigma_1 + \tau \beta S^z\,\,^{\rm *)}$ 
&
$\alpha =\dfrac{\sqrt{3}\varepsilon_{s}\Delta_{\rm SO}(V^{\pi}_{pp}-V^{\sigma}_{pp})}{18 (V_{sp})^2(R/a)}
\simeq \dfrac{-0.08\,{\rm meV}}{R[{\rm nm}]}\quad
\beta = \dfrac{-\sqrt{3}\Delta_{\rm SO}  V^{\pi}_{pp}\cos 3\theta }{3(V^{\pi}_{pp}+V^{\sigma}_{pp})(R/a)}
\simeq \dfrac{-0.31\, {\rm meV}}{R[{\rm nm}]} \cos 3 \theta$\\[10pt]\hline \\[-10pt]
$H_{\rm SO}^{\rm el}= \tau e E \xi S^y \sigma_2$ 
& 
$\xi = -\dfrac{\Delta_{\rm SO} }{3 V_{sp}} \xi_0 \simeq 2 \times 10^{-4} {\rm nm}\bs\bs\bs eE\xi\simeq0.2$ meV for $E=1$ V/nm\\
\hline\hline\\[-19pt]
\end{tabular}
\begin{flushleft}
$^{\rm *)}$ see also Refs. \onlinecite{ando,huertas-hernando,jeong,izumida}. 
\end{flushleft}
\vskip -28pt
\label{the_effective_hamiltonian}
\end{table*}

{\it Effective theory.} 
The microscopic model allows us to formulate an effective low-energy theory for the $\pi$ band near the Dirac points $\ve K$ and $\ve K'$. As explained, we include the curvature effects and the $s$-$p_r$ transition $H_E^{(2)}$ important for the SOI, and neglect the other inessential interactions. We have tested this against numerical solutions of the full Hamiltonian [Eq. (\ref{full_hamiltonian})]. We also checked that additional trigonometrically modulated perturbations, such as $s$-$p_t$ transitions or sublattice staggered potentials, do not change the spectrum qualitatively.

Hamiltonian (\ref{full_hamiltonian}) can be written as
$H = H_\pi + H_\sigma + H_{\pi\sigma} $,
where $H_{\pi}$ and $H_{\sigma}$ describe the $\pi$ and $\sigma$ bands, and 
$H_{\pi\sigma}$ the $\sigma\pi$ coupling. For momenta $k$ close to a Dirac point $||H_{\pi\sigma}|| \ll ||H_\pi - H_{\sigma}||$. In perturbation theory
we obtain $H_\pi^{\rm eff} = H_{\pi} + H_{\pi\sigma}[H_{\pi}-H_{\sigma}]^{-1} H_{\pi\sigma} + \mathcal O((a/R)^2)$. Here we keep only terms up to second order in the small parameter $a/R$ and the small energies $\Delta_{\rm SO},\; eE\xi_0$ which must be compared to typical hopping amplitudes $\sim$ eV. $H_{\pi\sigma} \simeq H^{\pi\sigma}_{\rm bs}+ H^{\pi\sigma}_{\rm SO}+H_E^{(2)}$,
where the superscript $\pi\sigma$ refers to the terms coupling the $\pi$ and the $\sigma$ bands. We calculate the effective Hamiltonian for the $\pi$ band
\begin{equation}
H_\pi^{\rm eff} = H_\pi^0 + H_{\rm orb}^{\rm cv} + H_{\rm SO}^{\rm cv} + H_{\rm SO}^{\rm el},
\end{equation}
where the last three terms are explicitly listed in Table \ref{the_effective_hamiltonian}, including numerical values for typical CNTs. Furthermore,  
$H_\pi^0 = \lim_{R\rightarrow\infty} H_\pi = \hbar v_F (k_G^0\sigma_1 + k \tau  \sigma_2)$ is the $\pi$ band Hamiltonian for flat graphene with periodic boundary conditions, with $\tau=\pm1$ labeling the two inequivalent $\ve K$ and $\ve K'$ points and $k$ the momentum along the tube measured from the corresponding Dirac point. For semiconducting CNTs, $k^0_G=(n-\tau\delta/3)/R\neq 0$ leads to a gap $2\hbar v_F| k_G^0|$, where $n\in \mathbb Z$ is the subband index and $\delta=(N_1-N_2){\rm mod}\,3$ for a $(N_1,N_2)$-CNT. In the following, we consider only the lowest subband in metallic CNTs defined by $k_G^0=0$ \cite{dresselhaus_book}.

$H_{\rm orb}^{\rm cv}$ describes the curvature induced $k$-shift of the Dirac points \cite{ando,izumida}, e.g., $\ve K\rightarrow \ve K-\Delta \ve k_{\rm cv}$, with $\Delta \ve k_{\rm cv} = (\Delta k_{\rm cv}^t , \Delta k_{\rm cv}^z)$. 
The shift $\Delta k_{\rm cv}^z$ is parallel to the tube and can be removed by a gauge transformation shifting the origin of $k$. For non-armchair CNTs, $\Delta k_{\rm cv}^t \neq 0$ and gaps are introduced by the curvature $H_{\rm orb}^{\rm cv}$.
$H_{\rm SO}^{\rm cv}$ contains the curvature induced SOI \cite{ando,huertas-hernando,izumida,jeong}. It contains only $S^z$
because $S^{r,t}$ depend on $\cos\varphi$ and $\sin\varphi$, which average out in the $\varphi$ integration.

On the other hand, $H_E^{(2)}\propto\cos\varphi$ which, in combination with the SOI terms involving $S^t = S^y\cos\varphi - S^x \sin\varphi$, leads to a nonvanishing 
$H_{\rm SO}^{\rm el} \propto S^y\sigma_2\, \Delta_{\rm SO}  eE\xi_0 \int d\varphi \cos^2\varphi$ \cite{efield_footnote}. Since the term proportional to $S^r$ couples only within the $\sigma$ band, it leads to negligible higher order corrections.
Hence,
\begin{equation}
H_{\rm SO}^{\rm el}= \tau e E \xi S^y \sigma_2,
\end{equation}
where $\xi = -\xi_0\Delta_{\rm SO}  / 3 V_{sp} $. This is one of our main results.

\begin{figure}[!htp]
\includegraphics[width=8.2cm]{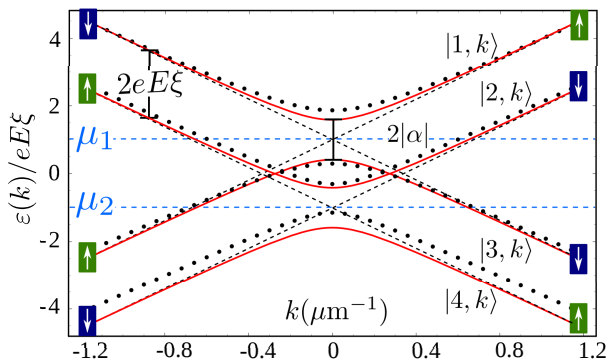}
\caption{(Color online) Energy dispersion $\varepsilon(k)$ of a (10,10) armchair nanotube (implying $R=0.67$nm). The solid (red) lines show the analytical results [see Eq. (\ref{analytical_spectrum}) and Tab. \ref{the_effective_hamiltonian}] and the dots show numerical results obtained from $H$ in Eq. (\ref{full_hamiltonian}). The axial $k$-shift $\Delta k_{\rm cv}^z$ has been removed in both, the numerical and the analytical spectrum. The arrows correspond to the $S^y$ projections and are reversed for $E \to -E$. The field strength is $E=1 \frac{\rm V}{\rm nm}$ so that the splitting $2eE\xi\simeq0.4$meV and the gap $2 |\alpha|\simeq0.24$meV. The dashed lines indicate the spectrum for the case $\alpha=0$ with the spin-degeneracies at $k=0$. The dashed gray (light blue) lines indicate chemical potentials at which only helical modes exist [see Eq. (\ref{Sy_k})].}
\vskip-7pt
\label{fig_spectrum}
\end{figure}

{\it Spectrum and helical states.} First we focus on armchair CNTs, assuming that $\Delta k^z_{\rm cv}$ has been gauged away. Furthermore, $\Delta k^t_{\rm cv}=0$ and $\beta=0$ so that the physics is completely determined by the interplay of $H_{\rm SO}^{\rm el}$ and $\alpha S^z\sigma_1$. In Fig. \ref{fig_spectrum}, we show the spectrum for a (10,10)-CNT in an electric field of 1V/nm. For $|k|\gg|\alpha/\hbar v_F|$, $H_{\rm SO}^{\rm el}$ aligns the spin in $y$-direction. For the right-moving branch ($\tau\sigma_2 = 1$, positive slope) the energy of the $S^y=\uparrow$ state is higher than the energy of the $S^y=\downarrow$ state by $2eE\xi$. For the left-moving branch ($\tau\sigma_2=-1$, negative slope) the $S^y = \downarrow$ state is higher in energy. 
Without the term $\alpha S^z\sigma_1$ the spectrum would be spin-degenerate at $k=0$ (dashed lines in Fig. \ref{fig_spectrum}). Unlike in usual one-dimensional conductors \cite{wires,spin_peierls} these degeneracies cannot be lifted by a uniform magnetic field because hybridization between the crossing bands requires the combination of spin flip and sublattice hybridization. This is, however, caused precisely by $\alpha S^z\sigma_1$, which is generated by virtual transitions to the $\sigma$ band that result in the simultaneous spin and sublattice hybridization. As a result, a gap of size 2$|\alpha|$ is opened at each degeneracy point. 
The resulting spectrum, shown in Fig. \ref{fig_spectrum}, at the $\ve K$ point has four branches, the subbands $\left|m,k\right>$, given by
\begin{equation}
\varepsilon(k) = \pm eE\xi \pm \sqrt{\alpha^2 + (\hbar v_F k)^2}. \label{analytical_spectrum}
\end{equation}
An equivalent spectrum exists at $\ve K'$. The spin orientations on the branches for $|k|\gg|\alpha/\hbar v_F|$ are identical at both Dirac points (arrows in Fig. \ref{fig_spectrum}). For general $ k$, the $S^y$ expectation value in state $\left|m,k\right>$ is given by
\begin{equation} \label{Sy_k}
\left<m,k\right|S^y\left|m,k\right> = \pm k / \sqrt{(\alpha/\hbar v_F)^2+{ k}^2},
\end{equation}
where for $eE\xi>0$ the $+$ corresponds to subbands $m=1,4$ in Fig. \ref{fig_spectrum}, and the $-$ to $m=2,3$ (and vice versa for $eE\xi<0$). Note also that the expectation values of $S^x$ and $S^z$ in all states give zero, so that only $\left<S^y\right>\neq 0$. In this sense, the states are always perfectly spin-polarized, even though the measured total spin is not unity. The bands crossing the chemical potentials $\mu_1$ and $\mu_2$ indicated in Fig. \ref{fig_spectrum} have $\left<S^y\right>\simeq \pm 0.95$.
We also note that electron-electron interactions generally lead to an enhancement of the gap $2|\alpha|$ \cite{spin_peierls}.

Fig. \ref{fig_spectrum} shows the analytical spectrum Eq. (\ref{analytical_spectrum}) for an armchair CNT in comparison with a numerical diagonalization of Hamiltonian (\ref{full_hamiltonian}). The qualitative features of the spectrum are well preserved by the effective theory.

\begin{figure}[!t]
\centering
\includegraphics[width=245pt]{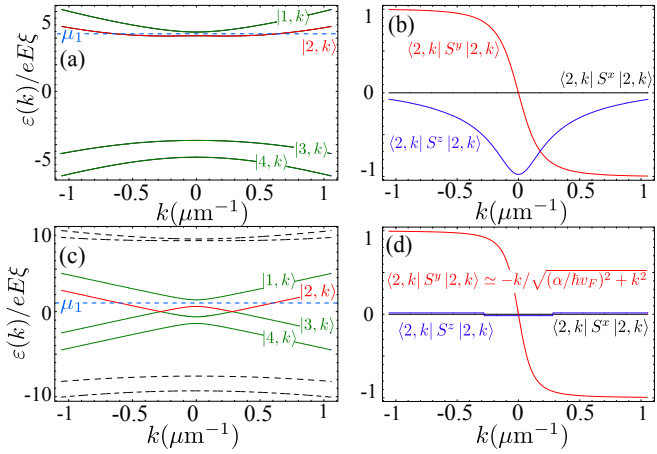}
\caption{(Color online) Chiral (17,14)-CNT with electric field $E=1$V/nm ($eE\xi \simeq 0.2$meV). 
(a) The spectrum and (b) the spin expectation values at the $\ve K/ \ve K'$ point for the $\left|2,k\right>$ subband and magnetic field $B_z=0$.
For $B_z=0.81$T, the bands (c) at $\ve K'$ (dashed lines) are gapped, while the spectrum at $\ve K$ (solid lines) has the same form as in the armchair case [Eq. (\ref{analytical_spectrum})]. The size of the gap is $2 |\alpha|=0.16$meV. 
 The spin expectation values (d) at $\ve K$ for $\left|2,k\right>$ follow closely the armchair case [see Eq. (\ref{Sy_k})].}
\label{fig_valley_suppression}
\end{figure}

If, in an armchair CNT, the chemical potential is tuned to $\mu_1$ or $\mu_2$ (see Fig. \ref{fig_spectrum}), the remaining conducting modes are helical, i.e., the direction of motion is coupled to the spin direction. In the present case, the spin points along $\ve E \times \ve v$, where $\ve v =\pm v_F \hat{\ve z}$ for right and left movers, respectively. In particular, this implies that $\ve E\rightarrow -\ve E$ also reverses the helicity, thus inverting the spin filtering. We note that the helical modes are stable against small deviations from the $(N,N)$-CNT (armchair) case with chiral angle $\theta=\frac{\pi}{6}$. The additional terms $\beta S^z$ and $\hbar v_F \Delta k_{\rm cv}^t \sigma_1$, which appear for $\theta \neq\frac{\pi}{6}$, partially align the spin in $z$-direction and open gaps at the zero-energy crossing points. We find that for metallic chiral CNTs, e.g. with $(N+3,N)$ and $N\simeq 10-20$, that are close to the armchair limit, good spin polarization ($\left<S^y\right>\sim90$\% and $\left<S^z\right><20$\%) can still be obtained (see also Fig. \ref{fig_valley_suppression}). 

{\it Valley suppression.} 
In chiral $(N+3l, N)$-CNTs, with $l=1,2,...$, it is possible to mostly restore the armchair spectrum and spinor properties for one Dirac point by the further application of a magnetic field $B_z$ along the tube.
As mentioned above, $\theta<\frac{\pi}{6}$ results in $\cos(3\theta)\neq 0$, thus leading to two additional terms in the Hamiltonian: a transverse $k$-shift $\hbar v_F \Delta k_{\rm cv}^t \sigma_1$ and an effective Zeeman field $\tau\beta S^z$. These terms have opposite signs at different Dirac points. The field $B_z$ leads to terms of the same form, yet with equal signs at both Dirac points, so that the chirality-induced $\cos(3\theta)$-terms can be canceled at one of the Dirac points, whereas they are doubled at the other.
Indeed, the orbital effect of $B_z$ adds $\Delta k^t_B = \pi B_z R/\Phi_0$ to $\Delta k^t_{\rm cv}$, with $\Phi_0$ the magnetic flux quantum. The Zeeman effect of $B_z$ adds $\mu_B B_z S^z$ to $\tau\beta S^z$. Due to the different radius ($R$) dependencies of the Zeeman and orbital terms, $R$ and $B_z$ can be chosen such that both $\cos(3\theta)$ terms in Table \ref{the_effective_hamiltonian} cancel at one Dirac point, provided that $\Delta_{\rm SO}>0$ \cite{foot_so_sign} (see Fig. \ref{fig_valley_suppression}). However, since $R$ cannot be chosen continuously, the cancellation is perfect only for one of the two terms. The Zeeman term can be removed at $\ve K$ with $B_z = -\mu_B/\beta$, but small gaps will remain at energy $\varepsilon=0$. On the other hand, if the $\varepsilon=0$ gaps are to be closed, $B_z$ must be tuned such that $\Delta k^t_{B} + \Delta k_{\rm cv}^t=0$. The small residual Zeeman term $\beta^* = \beta + \mu_B B_z$ ($|\beta^*| \ll |\beta|$) then leads to
\begin{equation}
\varepsilon = \pm \sqrt{(\beta^*)^2 + (eE\xi)^2} \pm \sqrt{(\hbar v_F k)^2+\alpha^2},
\end{equation}
and to a small spin-polarization $\left<S^z\right> \simeq \pm \beta^*/eE\xi$ ($\left<S^x\right>=0$ in all cases).

An illustrative example is the (17,14)-CNT with $B_z\simeq 0.81 $T, for which the orbital and Zeeman cancellations work particularly well. At $\ve K$ the spectrum and the spinor properties $\left<m,k\right|S^{x,y,z}\left|m,k\right>$ of an armchair CNT are restored, while at $\ve K'$ the curvature-induced gap $\hbar v_F \Delta k^t_{\rm cv}$ is amplified by a factor of 2 (see Fig. \ref{fig_valley_suppression}). This amplification is sufficient to remove all states of $\ve K'$ from the relevant energy range so that only $\ve K$ contributes a single pair of helical modes at the chemical potential $\mu_1$.

We note that as an immediate consequence of the SOI induced gaps the conductance of the CNT is reduced by a factor of two, and by an additional factor of two if the valley degeneracy is lifted.
As mentioned, helical modes can be used as spin filters, Cooper pair splitters, and allow for Majorana fermions at the CNT edges if the latter is brought in contact with a superconductor. These properties, together with the all-electric control, make CNTs attractive candidates for spintronic and quantum computing applications.

We acknowledge discussions with D. L. Maslov, and funding from the Swiss NSF, NCCR Nanoscience (Basel), and DARPA QuEST.


\begin{thebibliography}{0}
\bibitem{foot_so_sign} For $\Delta_{\rm SO}<0$, $\Delta k_{\rm cv}^t$ and $\tau\beta$ have opposite signs, so that either the transverse $k$-shift or the effective Zeeman field can be canceled by tuning $B_z$, but not both. 
However, a calculation based on hydrogenic orbitals and a $r^{-1}$ potential
shows that $\Delta_{\rm SO} > 0$.
\end{thebibliography}
\end{document}